\documentclass[letterpaper, 10 pt, conference]{ieeeconf}  

\IEEEoverridecommandlockouts                              
\usepackage{graphics} 
\usepackage{graphicx}
\usepackage{svg}
\usepackage[font=small]{caption}
\usepackage{subcaption}
\usepackage{epsfig} 
\usepackage{amsmath} 
\usepackage{amssymb}  

\usepackage{amsthm}
\usepackage{mathtools}
\usepackage{amsmath}
\usepackage{cite}
\usepackage{eucal}

\DeclareMathOperator*{\argminB}{argmin}
\usepackage{hyperref}
\let\labelindent\relax
\usepackage{enumitem}
\usepackage{stfloats} 

\newtheorem{definition}{Definition}
\newtheorem{problem}{Problem}
\newtheorem{theorem}{Theorem}
\newtheorem{lemma}{Lemma}
\newtheorem{remark}{Remark}
\newtheorem{assumption}{Assumption}

\newtheorem{proposition}{Proposition}
\usepackage{mathtools, cuted}

\usepackage{ulem} 
\usepackage{comment}
\usepackage{subcaption}
\usepackage{siunitx} 
\newcommand\red[1]{\textcolor{red}{#1}} 

\usepackage{tikz}

\newcommand\copyrighttext{%
  \footnotesize \textcopyright 2023 IEEE.  Personal use of this material is permitted.  Permission from IEEE must be obtained for all other uses, in any current or future media, including reprinting/republishing this material for advertising or promotional purposes, creating new collective works, for resale or redistribution to servers or lists, or reuse of any copyrighted component of this work in other works.}
\newcommand\copyrightnotice{%
\begin{tikzpicture}[remember picture,overlay]
\node[anchor=south,yshift=10pt] at (current page.south) {\fbox{\parbox{\dimexpr\textwidth-\fboxsep-\fboxrule\relax}{\copyrighttext}}};
\end{tikzpicture}%
}

\title{\LARGE \bf
Safety-Critical Control under Multiple State and Input Constraints \newline
and Application to Fixed-Wing UAV
}

\author{Donggeon David Oh{${}^{*}$}, Dongjae Lee{${}^{*}$}, and H. Jin Kim
\thanks{* The first two authors contributed equally to this work.}%
\thanks{This work was supported by Unmanned Vehicles Core Technology Research and Development Program through the National Research Foundation of Korea(NRF) and Unmanned Vehicle Advanced Research Center(UVARC) funded by the Ministry of Science and ICT(NRF-2020M3C1C1A010864).}%
\thanks{D. D. Oh is with the Department of Mechanical and Aerospace Engineering, Seoul National University, Seoul 08826, South Korea {\tt\small donggeonoh1999@snu.ac.kr}}%
\thanks{D. Lee and H. Jin Kim are with the Department of Aerospace Engineering and the Automation and Systems Research Institute (ASRI), Seoul National University, Seoul 08826, South Korea {\tt\small ehdwo713@snu.ac.kr, hjinkim@snu.ac.kr}}%
}

\begin{document}

\maketitle
\copyrightnotice
\thispagestyle{empty}
\pagestyle{empty}

\begin{abstract}

This study presents a framework to guarantee safety for a class of second-order nonlinear systems under multiple state and input constraints. To facilitate real-world applications, a safety-critical controller must consider multiple constraints simultaneously, while being able to impose general forms of constraints designed for various tasks (e.g., obstacle avoidance).
With this in mind, we first devise a zeroing control barrier function (ZCBF) using a newly proposed nominal evading maneuver. By designing the nominal evading maneuver to 1) be continuously differentiable, 2) satisfy input constraints, and 3) be capable of handling other state constraints, we deduce an ultimate invariant set, a subset of the safe set that can be rendered forward invariant with admissible control inputs. Thanks to the development of the ultimate invariant set, we then propose a safety-critical controller, which is a computationally tractable one-step model predictive controller (MPC) with guaranteed recursive feasibility. We validate the proposed framework in simulation, where a fixed-wing UAV tracks a circular trajectory while satisfying multiple safety constraints including collision avoidance, bounds on flight speed and flight path angle, and input constraints.

\end{abstract}

\section{Introduction} \label{introduction}

Safety is an essential factor to be considered in controller design. 
Safety in general encompasses various keywords including collision avoidance, input saturation, or task-space constraints \cite{tee2010adaptive} which frequently appear in control/robotics applications. 
When a \textit{single} state constraint is imposed on the given system with actuation limits, one widely adopted method of safety-critical controller design is to find a zeroing control barrier function (ZCBF) by which a subset of the safe set, herein called the ZCBF set, is guaranteed to be rendered forward invariant by an admissible control law \cite{squires2022composition, breeden2021high, agrawal2021safe, cortez2020correct, wang2018permissive}. 

However, in many cases, \textit{multiple} state constraints are imposed simultaneously on a system under input constraints. 
The most common approach for guaranteeing safety in the presence of such multiple constraints has been designing multiple ZCBFs, each of which is induced by a single state constraint, and then applying all ZCBF constraints at once in a quadratic program (QP) \cite{rauscher2016constrained, nguyen20163d, cortez2019control}. 
A major limitation of this approach is that the feasibility of the QP cannot be guaranteed; unlike each of the ZCBF sets, the intersection of the ZCBF sets may not be rendered forward invariant by any admissible control law.

Recently, a few strategies that attempt to alleviate such issue of controller infeasibility were presented.
A QP with guaranteed feasibility that addresses multiple ZCBFs was formulated in \cite{xu2018constrained}, but the consideration of input constraints was left for future work.
In \cite{cortez2022robust}, a controller that handles multiple ZCBFs as well as input constraints was proposed. 
However, the authors only considered the case of non-overlapping ZCBFs (i.e., ZCBFs with non-intersecting set boundaries), in which only one ZCBF acts at a time. 
Multiple ZCBFs that together ensure forward invariance of a safe set were constructed in \cite{cortez2022safe}, but the method is not applicable to state constraints such as obstacle avoidance constraints that cannot be written in the form of box constraints. 
In \cite{breeden2022compositions}, a strategy for decoupling the design of multiple ZCBFs in the presence of input constraints was introduced, but it may result in an overly conservative viability domain (i.e., controlled invariant set) due to the idea of shrinking the set of available control inputs.
To sum up, the existing methodologies either are applicable only to special cases of safety constraints, or may lead to an overly conservative invariant set.

The ability to handle multiple safety constraints in various forms is crucial for a safety-critical controller when considering its application to a real-world dynamical system, for instance, a fixed-wing unmanned aerial vehicle (UAV).
In order to prevent overly aggressive maneuvers and aerodynamic stall which significantly deteriorate control performance, flight path angle and flight speed should be bounded \cite{khare2022predictive, bayen2007aircraft}. 
Furthermore, the UAV should avoid collision with the surrounding obstacles. 
Lastly, control inputs should be constrained as per the actuation limits.
Such constraints that naturally arise from the safety requirements of actual applications are unlikely to obey the non-overlapping assumption nor be represented in the form of box constraints. 


Therefore, in this study, we present a framework to guarantee safety for a class of second-order nonlinear systems under input constraints and multiple state constraints.
The presented method addresses two types of state constraints: one that can be formulated as a function of states with relative degree two, and box constraints that bound the states with relative degree one.
Since the suggested method is able to handle any form of constraint function with relative degree two, it is applicable for tasks with complex workspace constraints including obstacle avoidance.

Our main contribution is the construction of the \textit{ultimate invariant set}, which is a subset of the safe set that could be rendered forward invariant by an admissible control law.
We do this by developing the method presented in \cite{breeden2021high}, where a \textit{nominal evading maneuver} was utilized to derive a ZCBF from a single constraint function.
However, unlike the original method, we design a new continuously differentiable nominal evading maneuver that takes into account all the state constraints.
The proposed nominal evading maneuver is designed specifically to render the ultimate invariant set in a non-conservative way, while being able to handle overlapping constraint functions (i.e. constraint functions whose 0-sublevel sets have intersecting boundaries).
Then, we formulate a safety-critical one-step model predictive controller (MPC) with guaranteed recursive feasibility, which is suitable for real-time applications. 
The safety-critical controller is applied to a fixed-wing UAV, and we validate the proposed approach in simulation.

\section{Preliminaries} \label{preliminaries}
\subsection{Notations} \label{preliminaries_notations}
The class of $r$-times continuously differentiable functions is denoted $\CMcal{C}^r$. Let $\partial\CMcal{S}$ denote the boundary of a set $\CMcal{S}$, and $\emptyset$ represent the empty set. Given a matrix $W\in\mathbb{R}^{n\times n}$ and a vector $\mathbf{x}\in\mathbb{R}^n$, $\left \| \mathbf{x}  \right \|^2_W$ is equivalent to $\mathbf{x}^\top W\mathbf{x}$. 
$L_f h\left(\mathbf{x} \right )=\frac{\partial h}{\partial \mathbf{x}}f\left(\mathbf{x} \right )$ denotes the Lie derivative of a function $h:\mathbb{R}^n\rightarrow\mathbb{R}$ along $f:\mathbb{R}^n\rightarrow\mathbb{R}^{n\times m}$ at point $\mathbf{x}\in\mathbb{R}^n$. 
A function $\alpha:\mathbb{R}\rightarrow\mathbb{R}$ belongs to extended class-$\CMcal{K}_\infty$ if $\alpha$ 
is strictly increasing, $\alpha\left(0\right)=0$, $\lim_{r\rightarrow\infty}\alpha\left(r\right)=\infty$, and $\lim_{r\rightarrow-\infty}\alpha\left(r\right)=-\infty$. 
$\left[N\right]$ is equivalent to the set $\left\{1, 2, \cdots, N\right\}$ for an integer $N\geq1$.
The subscript $i$ denotes the $i^{\textup{th}}$ element of the related vector.

\subsection{Safety and Set Invariance} \label{preliminaries_set_invariance}
Consider a nonlinear control-affine system
\begin{equation}
    \dot{\mathbf{x}}=f\left(\mathbf{x} \right )+g\left(\mathbf{x} \right )\mathbf{u},
    \label{eq.affine_form}
\end{equation}
with state $\mathbf{x}\in\CMcal{X}\subset\mathbb{R}^n$ and control input $\mathbf{u}\in\CMcal{U}\subset\mathbb{R}^m$. $\CMcal{U}$ represents the set of admissible control inputs, and $f:\CMcal{X}\rightarrow\mathbb{R}^n$ and $g:\CMcal{X}\rightarrow\mathbb{R}^{n\times m}$ are locally Lipschitz.

Assume that a part of the state-space $\CMcal{X}$ should be avoided. We formulate such constraint using a continuously differentiable constraint function $h:\CMcal{X}\rightarrow\mathbb{R}$ that defines the safe set $\CMcal{S}$ as the 0-sublevel set of $h$:
\begin{equation}
    \CMcal{S}:=\left \{ \mathbf{x} \in \CMcal{X}\mid h\left(\mathbf{x} \right )\leq 0\right \}.
    \label{eq.sublevel_set}
\end{equation}
We aim to design a control law $\mathbf{u}:\CMcal{X}\rightarrow\CMcal{U}$ that is guaranteed to keep the system (\ref{eq.affine_form}) inside $\CMcal{S}$. In this regard, we formally define safety using the concept of forward invariance.

\begin{definition} \textup{\textbf{\cite{ames2016control}}}
\label{def.forward_invariant} 
Let $\mathbf{u}=\pi\left(\mathbf{x}\right)$ be a feedback control law that induces the closed loop dynamics $\dot{\mathbf{x}}=f_{cl}\left(\mathbf{x}\right):=f\left(\mathbf{x}\right)+g\left(\mathbf{x}\right)\pi\left(\mathbf{x}\right)$
which is assumed to be locally Lipschitz. The set $\CMcal{S}$ is \uline{forward invariant} if $\mathbf{x}\left(t\right)\in\CMcal{S}$ for all $\mathbf{x}\left(0\right)\in\CMcal{S}$ and $t\geq0$. The closed-loop system $\dot{\mathbf{x}}=f_{cl}\left(\mathbf{x}\right)$ is \uline{safe} with respect to the set $\CMcal{S}$ if $\CMcal{S}$ is forward invariant.
\end{definition}


A sufficient condition for the set $\CMcal{S}$ to be rendered forward invariant and thus achieve safety is $h$ being a zeroing control barrier function, defined as follows.

\begin{definition} \textup{\textbf{\cite{ames2016control}}}
\label{def.ZCBF}
Let $S\subset\CMcal{X}$ be defined as (\ref{eq.sublevel_set}) for a continuously differentiable function $h:\CMcal{X}\rightarrow\mathbb{R}$. $h$ is a \uline{zeroing control barrier function (ZCBF)} if there exists an extended class-$\CMcal{K}_\infty$ function $\alpha$ such that for the control system (\ref{eq.affine_form}) and for all $\mathbf{x}\in\CMcal{X}$:
\begin{equation}
    \inf_{\mathbf{u}\in\CMcal{U}}\left [ L_fh\left ( \mathbf{x} \right )+L_gh\left(\mathbf{x} \right)\mathbf{u}-\alpha\left(-h\left(\mathbf{x} \right ) \right ) \right ]\leq0.
    \label{eq.ZCBF_condition}
\end{equation}
\end{definition}

\begin{theorem} \textup{\textbf{\cite{ames2016control, breeden2021high}}}
\label{thm.ZCBF}
Let $S\subset\CMcal{X}$ be defined as (\ref{eq.sublevel_set}) for a continuously differentiable function $h:\CMcal{X}\rightarrow\mathbb{R}$. If $h:\CMcal{X}\rightarrow\mathbb{R}$ is a ZCBF on $\CMcal{X}$ and $\frac{\partial h}{\partial\mathbf{x}}\left(\mathbf{x}\right)\neq\mathbf{0}$ for all $\mathbf{x}\in\partial\CMcal{S}$, then any Lipschitz continuous controller $\mathbf{u}:\CMcal{X}\rightarrow\CMcal{U}$ that satisfies
\begin{equation}
    L_fh\left(\mathbf{x} \right )+L_gh \left(\mathbf{x} \right )\mathbf{u}\left(\mathbf{x}\right)\leq\alpha \left(-h \left(\mathbf{x} \right ) \right ), \forall\mathbf{x}\in\CMcal{S}
    \label{eq.ZCBF_input}
\end{equation}
renders $\CMcal{S}$ forward invariant.
\end{theorem}


If $h$ satisfies (\ref{eq.ZCBF_condition}) for some extended class-$\CMcal{K}_\infty$ function $\alpha$, then $h$ is a valid ZCBF, and its 0-sublevel set $\CMcal{S}$ is referred to as a \textit{ZCBF set}. In practice, a constraint function is usually not a valid ZCBF unless it is designed specifically in consideration of (\ref{eq.ZCBF_condition}). This is generally due to the following two issues \cite{ames2019control}:


\begin{enumerate}[label=\textbf{I\arabic*.}, ref=I\arabic*]
\item \label{I1} The set of admissible control inputs $\CMcal{U}$ restricts the available control actions.
\item \label{I2} If the constraint function $h$ has a relative degree of $r\geq2$ with respect to the system (\ref{eq.affine_form}), no control authority exists since (\ref{eq.ZCBF_condition}) simplifies to $L_fh\left(\mathbf{x}\right)-\alpha\left(-h\left(\mathbf{x}\right)\right)\leq0$.
\end{enumerate}
In both cases, an extended class-$\CMcal{K}_\infty$ function $\alpha$ that satisfies (\ref{eq.ZCBF_condition}) is unlikely to exist. 

One way of addressing the issues \ref{I1} and \ref{I2} is to design a predefined \textit{nominal evading maneuver} $\mathbf{u}^*:\CMcal{X}\rightarrow\CMcal{U}$ that attempts to drive the system towards the interior of $\CMcal{S}$ \cite{breeden2021high, squires2022composition}. 
The flow operator $\psi_h\left(t;\mathbf{x}, \mathbf{u}^*\right)$ represents the value $h\left(\mathbf{y}\left(t\right)\right)$ resulting from the initial value problem $\dot{\mathbf{y}}=f\left(\mathbf{y}\right)+g\left(\mathbf{y}\right)\mathbf{u}^*\left(\mathbf{y}\right)$, $\mathbf{y}\left(0\right)=\mathbf{x}$, and was used to define a new ZCBF candidate $H:\CMcal{X}\rightarrow\mathbb{R}$ and the corresponding 0-sublevel set $\CMcal{S}_H$ in \cite{breeden2021high}.

\begin{theorem} \textup{\textbf{\cite{breeden2021high}}}
\label{thm.H_ZCBF}
The function $H:\CMcal{X}\rightarrow\mathbb{R}$ defined as $H\left(\mathbf{x}\right):=\sup_{t\geq0}\psi_h\left(t;\mathbf{x}, \mathbf{u}^*\right)$
is a valid ZCBF, provided $\CMcal{S}_H\neq\emptyset$. 
\end{theorem}
Note that the ZCBF set $\CMcal{S}_H$ is a subset of the safe set $\CMcal{S}$. A means to calculate $\dot{H}$ for general systems was also provided in \cite{breeden2021high}, given $\mathbf{u}^*\in\CMcal{C}^1$. Then, by Theorem \ref{thm.H_ZCBF}, any locally Lipschitz controller $\mathbf{u}:\CMcal{X}\rightarrow\CMcal{U}$ that satisfies $\dot{H}\left(\mathbf{x}, \mathbf{u}\right)\leq\alpha\left(-H\left(\mathbf{x}\right)\right)$ for an arbitrary extended class-$\CMcal{K}_\infty$ function $\alpha$ would render $\CMcal{S}_H$ forward invariant.

The method of using a nominal evading maneuver $\mathbf{u}^*$ to define $H\left(\mathbf{x}\right)$ alleviates the issues \ref{I1} and \ref{I2}. This is because at least one admissible control law $\mathbf{u}\in\CMcal{U}$, namely the nominal evading maneuver $\mathbf{u}^*$, is assured to satisfy (\ref{eq.ZCBF_condition}) for any extended class-$K_\infty$ function $\alpha$, and $H$ has a relative degree of 1 with respect to the system (\ref{eq.affine_form}).

\section{Problem Formulation} \label{problem_formulation}
\subsection{Dynamical Model} \label{problem_formulation_dynamical}
We consider a class of second-order nonlinear systems of the form
\begin{equation}
    \dot{\mathbf{x}} = \left[\begin{matrix} \dot{\mathbf{r}} \\ \dot{\mathbf{v}} \end{matrix}\right] = \left[\begin{matrix} {f}_{\mathbf{r}}\left(\mathbf{x}\right) \\ {f}_{\mathbf{v}}\left(\mathbf{x}\right) \end{matrix}\right] + \left[\begin{matrix} O_{n\times m} \\ g(\mathbf{x}) \end{matrix}\right] \mathbf{u},
    \label{eq.general_system}
\end{equation}
with state $\mathbf{x}:=\left[\mathbf{r}^\top,\mathbf{v}^\top\right]^\top\in\CMcal{X}\subset\mathbb{R}^{n+m}$ and control input $\mathbf{u}\in\CMcal{U}\subset\mathbb{R}^m$. Functions ${f}_{\mathbf{r}}:\CMcal{X}\rightarrow\mathbb{R}^{n}$, ${f}_{\mathbf{v}}:\CMcal{X}\rightarrow\mathbb{R}^{m}$, and $g:\CMcal{X}\rightarrow\mathbb{R}^{m \times m}$ are at least $\CMcal{C}^2$, where $g\left(\mathbf{x}\right)$ is defined as $g\left(\mathbf{x} \right ):=\textup{diag}\begin{bmatrix}
    g_{v_1}\left(\textbf{x}\right), g_{v_2}\left(\textbf{x}\right), \cdots, g_{v_m}\left(\textbf{x}\right)
\end{bmatrix}$ with $g_{v_i}:\CMcal{X}\rightarrow\mathbb{R}$, $i\in\left[m\right]$. 
States of relative degree 2 along (\ref{eq.general_system}) are denoted $\mathbf{r}\in\CMcal{X}_{\mathbf{r}}\subset\mathbb{R}^n$, and $\mathbf{v}\in\CMcal{X}_{\mathbf{v}}\subset\mathbb{R}^m$ represents the states of relative degree 1, and $\CMcal{X}=\CMcal{X}_{\mathbf{r}}\times\CMcal{X}_{\mathbf{v}}$. We will refer to $r_i$, $i\in\left[n\right]$ as \textit{RD2 states}, and $v_j$, $j\in\left[m\right]$ as \textit{RD1 states}. 
Dynamical models of a wide range of systems including fixed-wing UAVs \cite{lee2011obstacle}, adaptive cruise control problems \cite{ames2016control, ames2019control}, and spacecrafts \cite{breeden2021high, breeden2021guaranteed, agrawal2021safe} could be formulated in the form of (\ref{eq.general_system}). 

\subsection{Constraints} \label{problem_formulation_constraints}
Three types of constraints are considered in this work: RD2 constraint, RD1 constraints, and input constraints. 

The \textit{RD2 constraint function} $h_{\mathbf{r}}:\CMcal{X}_{\mathbf{r}}\rightarrow\mathbb{R}$ is formulated as a function of the RD2 states $\mathbf{r}$, and assumed to be at least $\CMcal{C}^2$. The associated safe set is defined as the 0-sublevel set of $h_{\mathbf{r}}$:
\begin{equation}
    \CMcal{S}_{\mathbf{r}} := \left\{\mathbf{x}=\left[\mathbf{r}^\top, \mathbf{v}^\top\right]^\top\in\CMcal{X}\mid h_{\mathbf{r}}\left(\mathbf{r}\right)\leq0\right\}.
    \label{eq.RD2_safe_set}
\end{equation}
We will refer to $\CMcal{S}_{\mathbf{r}}$ as the \textit{RD2 safe set}. If a dynamical model of an Euler-Lagrange system could be represented in the form of (\ref{eq.general_system}), then $\mathbf{r}$ would be equivalent to the generalized coordinates. Therefore, an RD2 constraint can be applied ubiquitously for tasks that involve designing a safe set dictated by the position of a system, e.g., obstacle avoidance. 

An RD1 constraint bounds the corresponding RD1 state in the form of a box constraint. Without loss of generality, we assume that RD1 constraints are applied to $v_1, v_2, \cdots, v_{c}$, where $c\leq m$. For an RD1 state $v_i$, $i\in\left[c\right]$, an \textit{RD1 constraint function} is formulated as
\begin{equation}
    h_{v_i}\left(\mathbf{v}\right):=\left(v_i-\tfrac{v^{min}_i+v^{max}_i}{2}\right)^2-\left(\tfrac{v^{max}_i - v^{min}_i}{2}\right)^2,
    \label{eq.RD1_constraint_function}
\end{equation}
where $v^{min}_i<v^{max}_i$ for all $i\in\left[c\right]$. The \textit{RD1 safe set} is then defined as the 0-sublevel set of $h_{v_i}$:
\begin{equation}
    \CMcal{S}_{v_i}:=\left\{\mathbf{x}=\left[\mathbf{r}^\top,\mathbf{v}^\top\right]^\top\in\CMcal{X}\mid h_{v_i}\left(\mathbf{v}\right)\leq0\right\},
    \label{eq.RD1_safe_set}
\end{equation}
which is equivalent to $\left\{\mathbf{x}\in\CMcal{X}\mid v^{min}_i\leq v_i\leq v^{max}_i\right\}$. Such type of constraint is highly applicable for general automated systems \cite{cortez2022safe}. 

The \textit{safe set} is defined as the intersection of the RD2 safe set and all RD1 safe sets:
\begin{equation}
    \CMcal{S}=\CMcal{S}_{\mathbf{r}}\cap\left(\textstyle{\bigcap_{i=1}^{c}}\CMcal{S}_{v_i}\right).
    \label{eq.safe_set}
\end{equation}
Input constraints represent the limited actuation capabilities of real-world systems. The set of admissible control inputs is defined as follows:
\begin{equation}
    \CMcal{U}:=\left\{\mathbf{u}\in\mathbb{R}^m\mid u^{min}_i\leq u_i\leq u^{max}_i, \vspace{0.2cm}\forall i\in\left[m\right]\right\},
    \label{eq.input_constraint}
\end{equation}
where $u^{min}_i< u^{max}_i$ for all $i\in\left[m\right]$.
Therefore, the problem we aim to solve can be formally stated as follows: 
\begin{problem}
\label{Prob.1}
Given the system (\ref{eq.general_system}), find a subset of the safe set $\CMcal{S}$ (\ref{eq.safe_set}) that could be rendered forward invariant by an admissible control law $\mathbf{u}:\CMcal{X}\rightarrow\CMcal{U}$. Then, design a safety-critical controller that is always feasible if the system state is inside such subset.
\end{problem}

As mentioned in Section \ref{introduction}, there exist several methodologies that ensure forward invariance of a subset of the safe set by designing multiple ZCBFs.
However, since the boundary of the RD2 safe set cannot be represented using box constraints and clearly intersects with the boundaries of RD1 safe sets, the strategies from \cite{cortez2022safe, cortez2022robust} cannot be applied to the system of interest. Moreover, the input constraints preclude the application of the methodology presented in \cite{xu2018constrained}.
To this end, instead of constructing multiple ZCBFs, we propose a new methodology to obtain an invariant set. 


\section{Safety-Critical Controller Design} \label{safety-critical}
In this section, we first present issues that need to be addressed in the design of nominal evading maneuver. Then, a nominal evading maneuver which satisfies the input constraints and takes into account multiple state constraints is proposed. Using the nominal evading maneuver, an \textit{ultimate safe set}, a subset of the safe set which can be rendered forward invariant using admissible control inputs, is defined. Finally, we construct a safety-critical one-step MPC with guaranteed feasibility that utilizes the invariance of the ultimate safe set.

\subsection{Issues in Nominal Evading Maneuver Design}

Recall the RD2 constraint function $h_{\mathbf{r}}$ and the RD2 safe set $\CMcal{S}_{\mathbf{r}}$. The RD2 constraint function is not likely to be a valid ZCBF because of the issues \ref{I1} and \ref{I2}: the system is under input constraints (\ref{eq.input_constraint}), and the relative degree of $h_{\mathbf{r}}$ with respect to the system (\ref{eq.general_system}) is 2. The latter can be easily observed by computing derivatives of the RD2 constraint function $\dot{h}_{\mathbf{r}}$ and $\ddot{h}_{\mathbf{r}}$ as
\begin{equation*}
\begin{aligned}
    \dot{h}_{\mathbf{r}}\left(\mathbf{x}\right)&=L_fh_{\mathbf{r}}\left(\mathbf{x}\right)+L_gh_{\mathbf{r}}\left(\mathbf{x}\right)\mathbf{u}=L_fh_{\mathbf{r}}\left(\mathbf{x}\right), \\
    \ddot{h}_{\mathbf{r}}\left(\mathbf{x}, \mathbf{u}\right)
    &=L^2_fh_{\mathbf{r}}\left(\mathbf{x}\right)+\textstyle{\sum_{i=1}^{m}}d_i\left(\mathbf{x}\right)u_i,
\end{aligned}
\end{equation*}
where $d_i\left(\mathbf{x}\right):=\frac{\partial\dot{h}_\mathbf{r}}{\partial v_i}g_{v_i}\left(\mathbf{x}\right)$ for all $i\in\left[m\right]$. Therefore, we adopt the methodology presented in Theorem \ref{thm.H_ZCBF} and design a nominal evading maneuver $\mathbf{u}^*:\CMcal{X}\rightarrow\CMcal{U}$ that allows us to define a valid ZCBF $H_{\mathbf{r}}:\CMcal{X}\rightarrow\mathbb{R}$. We will refer to $H_{\mathbf{r}}$ as the \textit{RD2 ZCBF}. The corresponding \textit{RD2 ZCBF set} $\CMcal{S}_{H_{\mathbf{r}}}$ is a subset of $\CMcal{S}_{\mathbf{r}}$ and is rendered forward invariant by a control law that satisfies the input constraints \cite{breeden2021high}.

A nominal evading maneuver should be designed to effectively drive the system towards the interior of the RD2 ZCBF set $\CMcal{S}_{H_{\mathbf{r}}}$. However, since the relative degree of the RD2 constraint function is 2, both $h_{\mathbf{r}}$ and $\dot{h}_{\mathbf{r}}$ cannot be manipulated directly with the control input $\mathbf{u}$. Thus, one alternative approach for designing a greedy nominal evading maneuver $\mathbf{u}_{\textup{greedy}}^*$ would be pointwise minimizing $\ddot{h}_{\mathbf{r}}$ as
\begin{equation}
    \mathbf{u}^*_{\textup{greedy}}\left(\mathbf{x}\right):=\argminB_{\mathbf{u}\in\CMcal{U}}\textstyle{\sum_{i=1}^{m}}d_i\left(\mathbf{x}\right)u_i.
    \label{eq.u_ball}
\end{equation}
Such nominal evading maneuver was shown to be effective for the task of safety-critical obstacle avoidance of a spacecraft in \cite{breeden2021high}.

Unfortunately, $\mathbf{u}^*_{\textup{greedy}}$ cannot be applied to the system (\ref{eq.general_system}) under the presence of RD1 constraints because of the following issues:
\begin{enumerate} [label=\textbf{I\arabic*.}, ref=I\arabic*]
    \setcounter{enumi}{2}
    \item \label{I3} Under the input constraints given as (\ref{eq.input_constraint}), $\mathbf{u}^*_{\textup{greedy}}$ from (\ref{eq.u_ball}) is equivalent to
    \begin{equation}
    u^*_{\textup{greedy}, i}\left(\mathbf{x}\right)=\left\{\begin{matrix}
    u^{min}_i \hspace{0.3cm}\textup{if}\hspace{0.3cm}d_i\left(\mathbf{x} \right )>0\\
    u^{max}_i \hspace{0.3cm}\textup{if}\hspace{0.3cm}d_i\left(\mathbf{x} \right )<0
    \end{matrix}\right.
    \label{eq.u_ball_i}
    \end{equation}
    for all $i\in\left[m\right]$. Therefore $\mathbf{u}^*_{\textup{greedy}}\left(\mathbf{x}\right)\notin\CMcal{C}^1$, and $\dot{H}$ cannot be calculated using the methodology presented in \cite{breeden2021high}. This is problematic because $\dot{H}$ is required to impose the ZCBF constraint in the form of (\ref{eq.ZCBF_input}).
    \item \label{I4} $\mathbf{u}^*_{\textup{greedy}}\left(\mathbf{x}\right)$ does not take into account the RD1 constraints. For example, if $v_i=v^{min}_i$, $d_i\left(\mathbf{x}\right)>0$, and $g_{v_i}\left(\mathbf{x}\right)>0$ for some $i\in\left[c\right]$, then from Assumption \ref{ass.2} which will be presented shortly afterwards, $u^*_{\textup{greedy}, i}\left(\mathbf{x}\right)$ renders $\dot{v}_i<0$. That is, $\mathbf{u}^*_{\textup{greedy}}$ cannot constrain the RD1 states to be inside the corresponding RD1 safe sets.
\end{enumerate}

To address the issues \ref{I3} and \ref{I4}, in the next subsection, we propose a new nominal evading maneuver $\mathbf{u}^*:\CMcal{X}\rightarrow\CMcal{U}$ that attempts to drive the system (\ref{eq.general_system}) towards the interior of $\CMcal{S}_{H_{\mathbf{r}}}$, while being at least $\CMcal{C}^1$ and handling the RD1 constraints. Before entering the next subsection, we introduce a modified input $\tilde{\mathbf{u}}:\CMcal{X}\times\CMcal{U}\rightarrow\mathbb{R}^m$ to consider nonzero $f_{\mathbf{v}}(\mathbf{x})$ term in (\ref{eq.general_system}) and possibly asymmetric input constraints (i.e., $-u^{min}_i \neq u^{max}_i$).

The modified input is defined in an elementwise manner:
\begin{equation}
    \tilde{u}_i\left(\mathbf{x}, \mathbf{u}\right):=\tfrac{f_{v_i}\left(\mathbf{x}\right)}{g_{v_i}\left(\mathbf{x}\right)}+u_i,
\label{eq.modified_input}
\end{equation}
for all $i\in\left[m\right]$.
Then, derivatives of the RD1 states are computed as
\begin{equation}    \dot{v}_i=f_{v_i}\left(\mathbf{x}\right)+g_{v_i}\left(\mathbf{x}\right)u_i=g_{v_i}\left(\mathbf{x}\right)\tilde{u}_i\left(\mathbf{x}, \mathbf{u}\right)
    \label{eq.v_i_dot}
\end{equation}
for all $i\in\left[m\right]$.
We see that $\dot{v}_i$ is solely dependent on a single input channel $u_i$. For the ease of controller design, we define functions $\mu_i:\CMcal{X}\rightarrow\mathbb{R}$ and $\nu_i:\CMcal{X}\rightarrow\mathbb{R}$ as
\begin{equation}
    \mu_i\left(\mathbf{x}\right):=-u^{min}_i-\tfrac{f_{v_i}\left(\mathbf{x}\right)}{g_{v_i}\left(\mathbf{x}\right)}\hspace{0.1cm},\hspace{0.2cm}\nu_i\left(\mathbf{x}\right):=u^{max}_i+\tfrac{f_{v_i}\left(\mathbf{x}\right)}{g_{v_i}\left(\mathbf{x}\right)}
    \label{eq.mu_i_nu_i}
\end{equation}
for all $i\in\left[m\right]$, where $-\mu_i\left(\mathbf{x}\right)$ and $\nu_i\left(\mathbf{x}\right)$ represent the minimum and maximum admissible values of the modified input $\tilde{u}_i\left(\mathbf{x}, \mathbf{u}\right)$ at $\mathbf{x}\in\CMcal{X}$.

We now state the two assumptions that are required to further the discussion of safety-critical controller design.
\begin{assumption}
\label{ass.1}
$g_{v_i}\left(\mathbf{x}\right)\neq0$ for all $i\in\left[m\right]$ and $\mathbf{x}\in\CMcal{X}$. 
\end{assumption}
\begin{assumption}
\label{ass.2}
$-\mu_i(\mathbf{x}) <0< \nu_i(\mathbf{x})$ for all $i\in\left[m\right]$ and $\mathbf{x}\in\CMcal{X}$.
\end{assumption}

The two assumptions are essential to ensure that the system (\ref{eq.general_system}) maintains sufficient control authority and are also widely underlain in other existing works \cite{cortez2022safe,breeden2021high}. If either of the two assumptions is violated, then the system loses control authority for $v_i$ or the sign of $\dot{v}_i = g_{v_i}(\mathbf{x}) (u_i + \frac{f_{v_i}\left(\mathbf{x}\right)}{g_{v_i}\left(\mathbf{x}\right)})$ becomes uncontrollable at some $\mathbf{x}\in\CMcal{X}$.
Therefore, Assumptions \ref{ass.1} and \ref{ass.2} are essential for controlling the given system (\ref{eq.general_system}), let alone guaranteeing its safety. 

Next, to consider input constraints for the modified input $\tilde{u}$, for all $i\in\left[m\right]$, we define $\tilde{u}^{max}_i:\CMcal{X}\rightarrow\mathbb{R}$ as a smooth approximation of $\min\left(\mu_i\left(\mathbf{x}\right), \nu_i\left(\mathbf{x}\right)\right)$:
\begin{equation}
    \tilde{u}^{max}_i\left(\mathbf{x}\right):= \tfrac{\mu_i\left(\mathbf{x}\right)+\nu_i\left(\mathbf{x}\right)-\sqrt{\left(\mu_i\left(\mathbf{x}\right)-\nu_i\left(\mathbf{x}\right)\right)^2+\epsilon}}{2},
    \label{eq.u_tilde_i_max}
\end{equation}
where $\epsilon\in\mathbb{R}_+$ is a small scalar for numerical stability.
Such smooth approximation of the $\min\left(\cdot,\cdot\right)$ operator is adopted in order to design a continuously differentiable nominal evading maneuver and therefore resolve the issue \ref{I3}.
From Assumption \ref{ass.2}, by taking sufficiently small $\epsilon$ that satisfies $\epsilon<4\mu_i\left(\mathbf{x}\right)\nu_i\left(\mathbf{x}\right)$ for all $i\in\left[m\right]$ and $\mathbf{x}\in\CMcal{X}$, $\tilde{u}^{max}_i\left(\mathbf{x}\right)$ is assured to be greater than 0.
Any modified input $\tilde{u}_i\left(\mathbf{x}, \mathbf{u}\right)$ that satisfies $-\tilde{u}^{max}_i\left(\mathbf{x}\right)\leq\tilde{u}_i\left(\mathbf{x}, \mathbf{u}\right)\leq\tilde{u}^{max}_i\left(\mathbf{x}\right)$ is admissible with respect to the input constraints (\ref{eq.input_constraint}). We utilize this property later in Lemma \ref{lem.nominal_in_admissible}.

\subsection{Nominal Evading Maneuver Design} \label{safety-critical_nominal}
We first design a nominal evading maneuver in terms of the modified input $\tilde{u}^*_i:\CMcal{X}\rightarrow\mathbb{R}$ for all $i\in\left[m\right]\setminus\left[c\right]$. 
In this case, we need not worry about the issue \ref{I4} because for all $i\in\left[m\right]\setminus\left[c\right]$, the RD1 state $v_i$, which is dictated by $\tilde{u}_i$, is not constrained.
However, we do need to make sure that $\tilde{u}^*_i$ is at least $\CMcal{C}^1$ to alleviate the issue \ref{I3}.
In this regard, we propose the following nominal evading maneuver in terms of the modified input $\tilde{u}^*_i$ for all $i\in\left[m\right]\setminus\left[c\right]$ that mimics the greedy control law $\mathbf{u}^*_{\textup{greedy}}$ from (\ref{eq.u_ball}):
\begin{equation}
    \tilde{u}^*_i\left(\mathbf{x}\right):=\tilde{u}^{max}_i\left(\mathbf{x}\right)\tanh\left(- k_id_i\left(\mathbf{x}\right)\right),
    \label{eq.u_tilde_not_RD1}
\end{equation}
where $k_i\in\mathbb{R}_+$ is a control gain. 
$\tilde{u}^*_i\left(\mathbf{x}\right)$ quickly approaches $-\tilde{u}^{max}_i\left(\mathbf{x}\right)$ as $d_i\left(\mathbf{x}\right)$ increases from zero, and converges to $\tilde{u}^{max}_i\left(\mathbf{x}\right)$ as $d_i\left(\mathbf{x}\right)$ decreases from zero. 
Such behavior is similar to that of $u^*_{\textup{greedy}, i}$ from (\ref{eq.u_ball_i}) with the difference being that $\tilde{u}^*_i$ is continuously differentiable, while $u^*_{\textup{greedy}, i}$ is not.

On the contrary, for all $i\in\left[c\right]$, the RD1 state $v_i$ should be constrained inside the RD1 safe set $\CMcal{S}_{v_i}$ defined as (\ref{eq.RD1_safe_set}). 
Therefore, for all $i\in\left[c\right]$, the nominal evading maneuver in terms of the modified input $\tilde{u}^*_i:\CMcal{X}\rightarrow\mathbb{R}$ should attempt to drive the system towards the interior of $\CMcal{S}_{H_{\mathbf{r}}}$, while addressing the issues \ref{I3} and \ref{I4}.
We present such nominal evading maneuver $\tilde{u}^*_i$ for all $i\in\left[c\right]$ in (\ref{eq.u_tilde_RD1}), where $v^d_i={\left(v^{max}_i-v^{min}_i\right)}/{2}$, $v^s_i={\left(v^{max}_i+v^{min}_i\right)}/{2}$, and $k_{i, 1}, k_{i, 2}, k_{i, 3}\in\mathbb{R}_+$ are control gains.

\begin{figure*}[t!]
\begin{equation}
    \tilde{u}^*_i\left(\mathbf{x}\right):=\tilde{u}^{max}_i\left(\mathbf{x}\right)\tanh\left(-k_{i, 1}g_{v_i}\left(\mathbf{x}\right)\right)\tanh\left(k_{i, 2}\left(v_i-\left(v^d_i\tanh\left(-k_{i, 3}g_{v_i}\left(\mathbf{x}\right)d_i\left(\mathbf{x}\right)\right)+v^s_i\right)\right)\right)
    \label{eq.u_tilde_RD1}
\end{equation}
\hrule
\vspace{-0.5cm}
\end{figure*}


It can be seen from (\ref{eq.u_tilde_RD1}) that the nominal evading maneuver in terms of the modified input $\tilde{u}^*_i$ is continuously differentiable for all $i\in\left[c\right]$, thereby alleviating the issue \ref{I3}. Moreover, $\tilde{u}^*_i$ attempts to effectively drive the system towards the interior of the RD2 ZCBF set, while resolving the issue \ref{I4} by guaranteeing $v_i$ to remain inside the RD1 safe set $\CMcal{S}_{v_i}$ for all $i\in\left[c\right]$. We first provide a brief explanation of such property using plots of $\tilde{u}^*_i$ against $v_i$, and then present a formal proof in the next subsection.

\begin{figure}[t!]
     \centering
     \begin{subfigure}[b]{4.25cm}
         \centering
         \includegraphics[width=\textwidth]{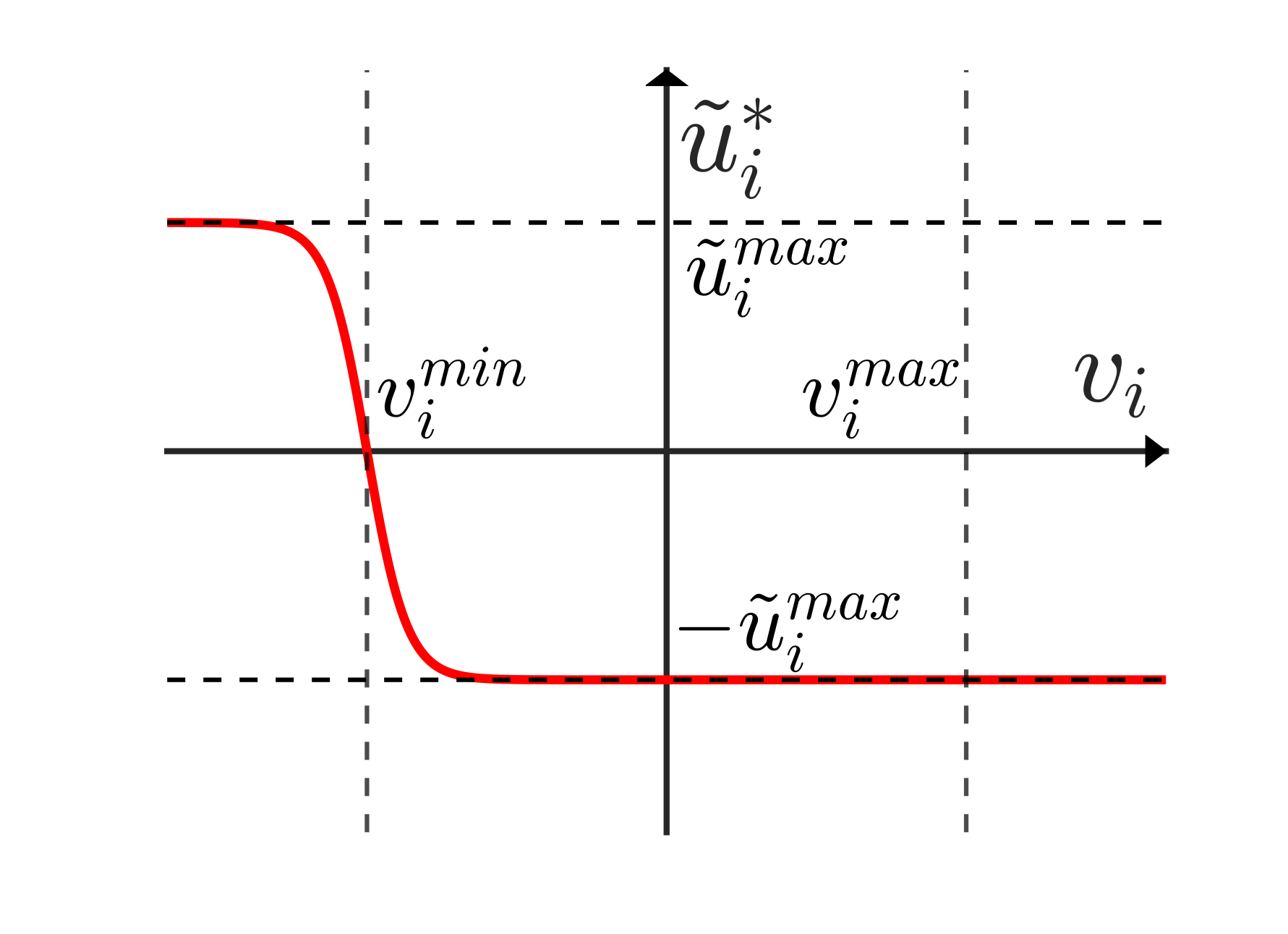}
         \caption{$d_i\left(\mathbf{x}\right)>0$, $g_{v_i}\left(\mathbf{x}\right)>0$}
         \label{fig.d>0g>0}
     \end{subfigure}
     \begin{subfigure}[b]{4.25cm}
         \centering
         \includegraphics[width=\textwidth]{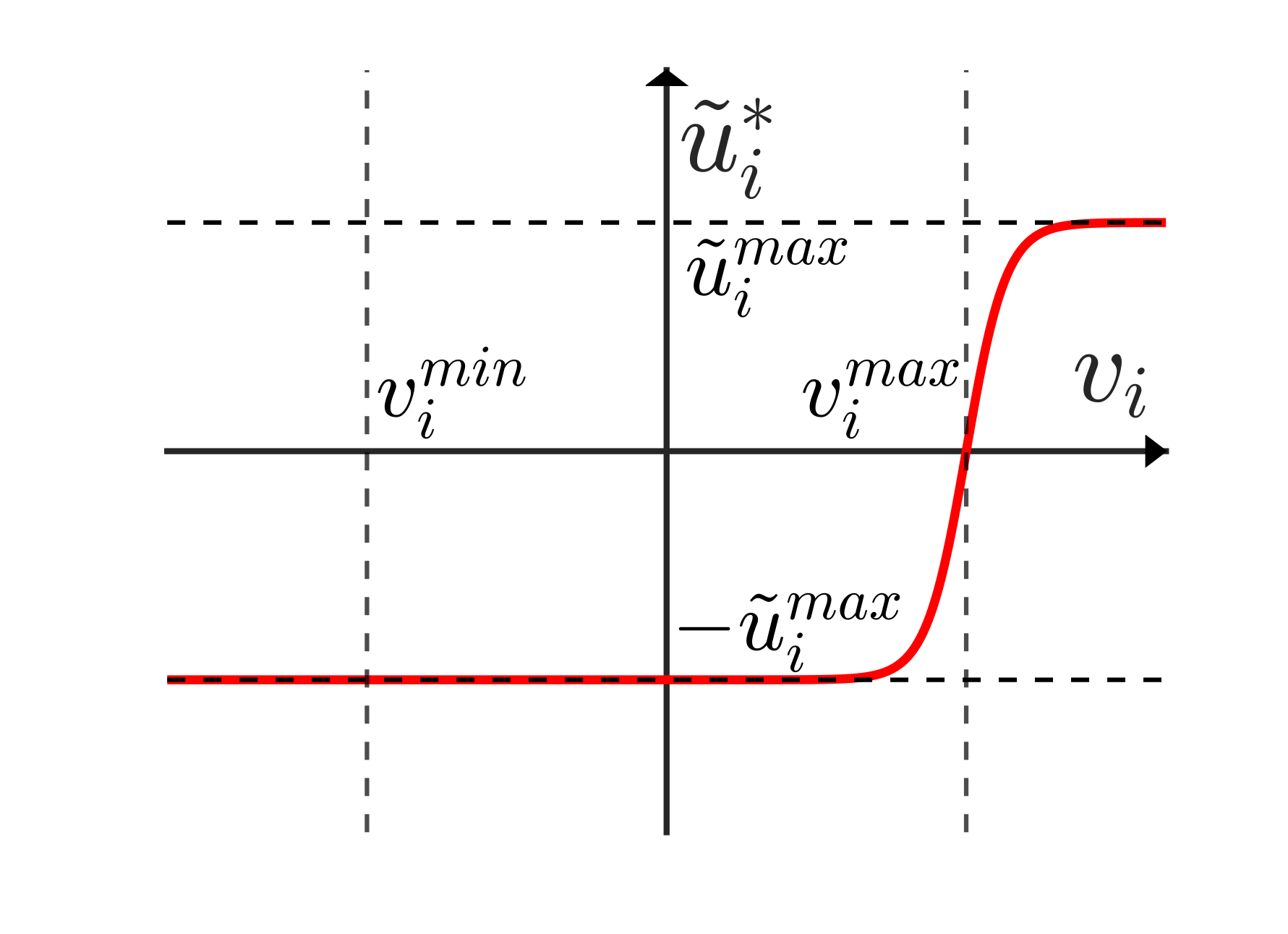}
         \caption{$d_i\left(\mathbf{x}\right)>0$, $g_{v_i}\left(\mathbf{x}\right)<0$}
         \label{fig.d>0g<0}
     \end{subfigure}
     \begin{subfigure}[b]{4.25cm}
         \centering
         \includegraphics[width=\textwidth]{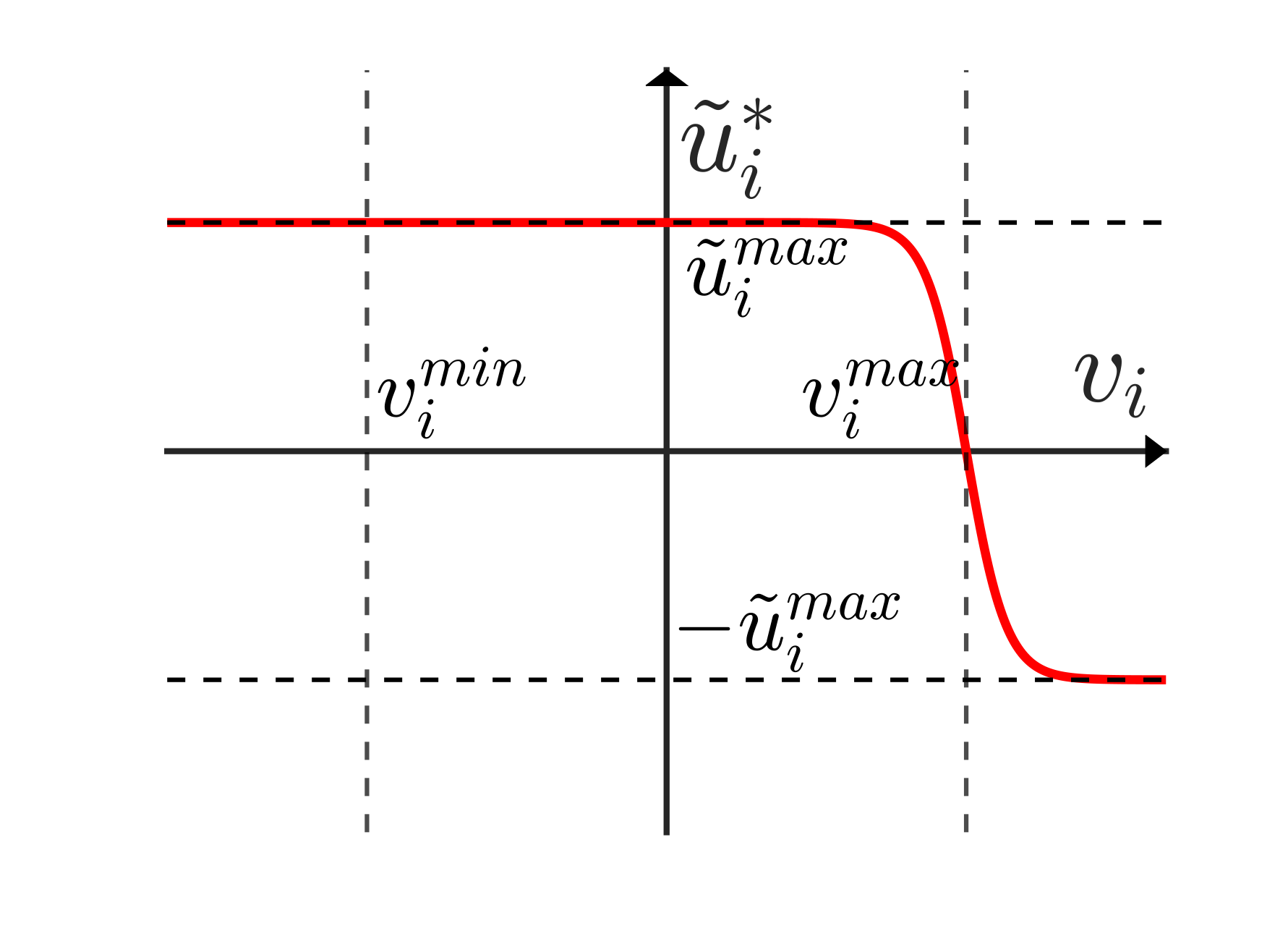}
         \caption{$d_i\left(\mathbf{x}\right)<0$, $g_{v_i}\left(\mathbf{x}\right)>0$}
         \label{fig.d<0g>0}
     \end{subfigure}
     \begin{subfigure}[b]{4.25cm}
         \centering
         \includegraphics[width=\textwidth]{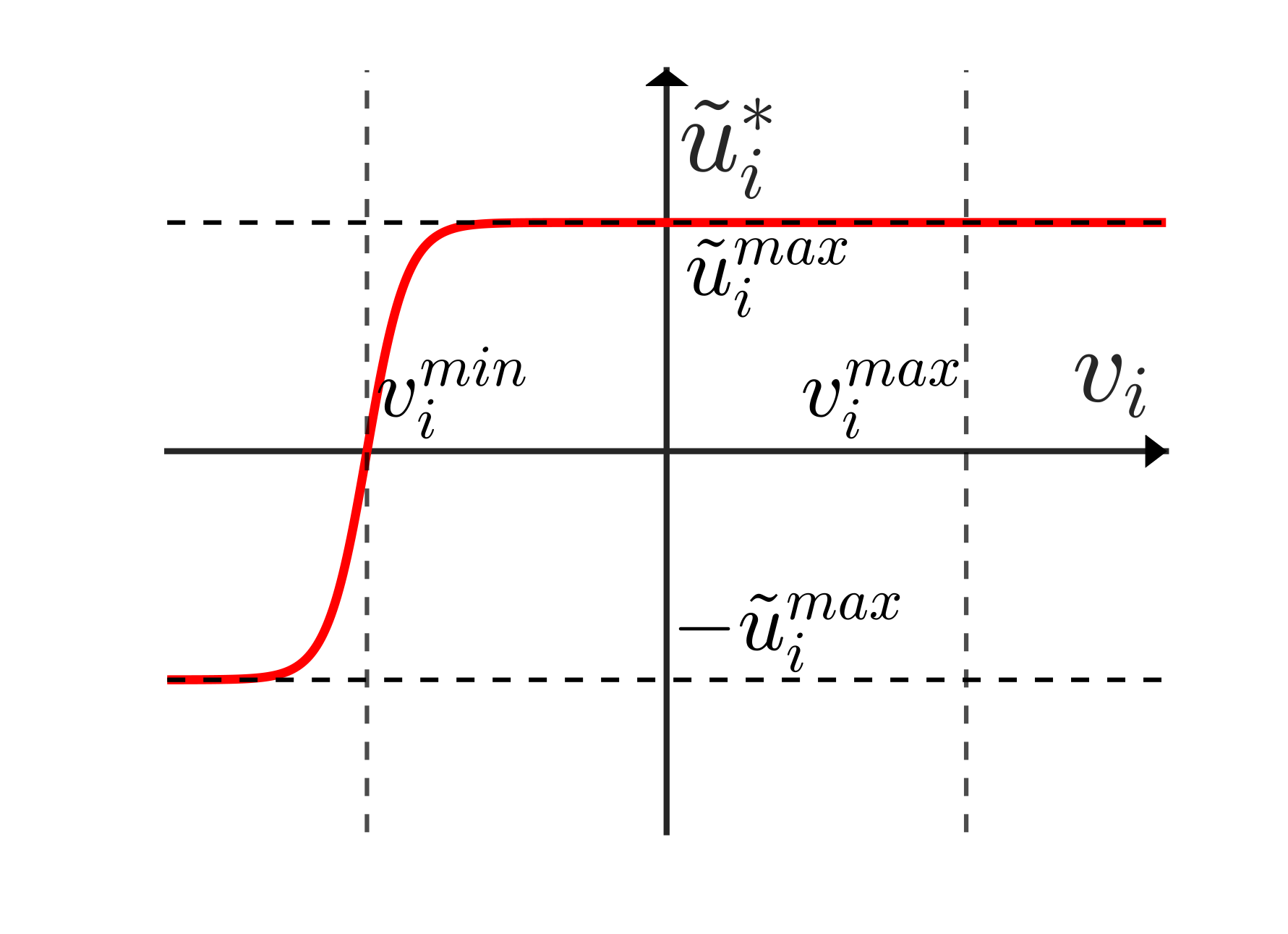}
         \caption{$d_i\left(\mathbf{x}\right)<0$, $g_{v_i}\left(\mathbf{x}\right)<0$}
         \label{fig.d<0g<0}
     \end{subfigure}
        \caption{Plots of the nominal evading maneuver in terms of the modified input $\tilde{u}^*_i\left(\mathbf{x}\right)$ against $v_i$ for $i\in\left[c\right]$ given different signs of $d_i\left(\mathbf{x}\right)$ and $g_{v_i}\left(\mathbf{x}\right)$. $\tilde{u}^*_i\left(\mathbf{x}\right)$ pointwise minimizes $\ddot{h}_{\mathbf{r}}\left(\mathbf{x}, \mathbf{u}\right)$ while assuring $v_i$ to remain inside the RD1 safe set $\CMcal{S}_{v_i}$.}
        \label{fig.modified_input}
        \vspace{-0.5cm}
\end{figure}

Plot of the nominal evading maneuver in terms of the modified input $\tilde{u}^*_i\left(\mathbf{x}\right)$ against $v_i$ given $d_i\left(\mathbf{x}\right)>0$ and $g_{v_i}\left(\mathbf{x}\right)>0$ is shown in Fig. \ref{fig.d>0g>0}.
The nominal evading maneuver is designed to satisfy $\tilde{u}^*_i\left(\mathbf{x}\right)\geq0$ if $v_i\leq v^{min}_i$, resulting in $\dot{v}_i\geq0$ from (\ref{eq.v_i_dot}).
As $v_i$ increases from $v^{min}_i$, $\tilde{u}^*_i\left(\mathbf{x}\right)$ quickly converges to $-\tilde{u}^{max}_i\left(\mathbf{x}\right)$.
Similar analyses can be done to Figs. \ref{fig.d>0g<0}, \ref{fig.d<0g>0} and \ref{fig.d<0g<0}. 
Taken together, $\tilde{u}^*_i\left(\mathbf{x}\right)$ attempts to pointwise minimize $\ddot{h}_{\mathbf{r}}$ in a similar manner as the greedy control law $u^*_{\textup{greedy}, i}\left(\mathbf{x}\right)$ from (\ref{eq.u_ball_i}). However, unlike $u^*_{\textup{greedy}, i}\left(\mathbf{x}\right)$, $\tilde{u}^*_i\left(\mathbf{x}\right)$ is able to guarantee $v_i$ to remain inside the corresponding RD1 safe set $S_{v_i}$. This will be formally proven in Theorem \ref{thm.main}.

We now reformulate the nominal evading maneuver in terms of the original control input. Using the definition of the modified input from (\ref{eq.modified_input}) and $\tilde{u}^*_i$ defined as (\ref{eq.u_tilde_not_RD1}) for $i\in\left[m\right]\setminus\left[c\right]$ and (\ref{eq.u_tilde_RD1}) for $i\in\left[c\right]$, the nominal evading maneuver in terms of the original input $\mathbf{u}^*:\CMcal{X}\rightarrow\CMcal{U}$ is derived as follows:
\begin{equation}
    u^*_i\left(\mathbf{x}\right) = \tilde{u}^*_i\left(\mathbf{x}\right)-\tfrac{f_{v_i}\left(\mathbf{x}\right)}{g_{v_i}\left(\mathbf{x}\right)}, \forall i\in\left[m\right].
    \label{eq.nominal_evading_maneuver}
\end{equation}
We restate the assessment of Fig. \ref{fig.modified_input} in the following lemma.
\begin{lemma}
\label{lem.RD1_safe_set_forward_invariance}
The nominal evading maneuver $\mathbf{u}^*$ from (\ref{eq.nominal_evading_maneuver}) renders the derivative of an RD1 constraint function nonpositive, i.e., $\dot{h}_{v_i}\left(\mathbf{x}, \mathbf{u}^*\right)\leq0$, for all $\mathbf{x}\in\partial\CMcal{S}_{v_i}$, $i\in\left[c\right]$.
\end{lemma}
\noindent \textit{proof.}
For all $i\in\left[c\right]$, the derivative of $h_{v_i}\left(\mathbf{v}\right)$ under the system dynamics (\ref{eq.general_system}) and the control law $\mathbf{u}^*$ is given as
\begin{equation}
    \label{eq.h_v_i_dot}
    \dot{h}_{v_i}\left(\mathbf{x}, \mathbf{u}^*\right)=2\left(v_i-\tfrac{v^{min}_i+v^{max}_i}{2}\right)g_{v_i}\left(\mathbf{x}\right)\tilde{u}^*_i\left(\mathbf{x}\right),
\end{equation}
where $\tilde{u}^*_i\left(\mathbf{x}\right)$ is the nominal evading maneuver in terms of the modified input from (\ref{eq.u_tilde_RD1}). 

First, we consider the case where $v_i=v^{max}_i$. Since $\tanh\left(-k_{i, 3}g_{v_i}\left(\mathbf{x}\right)d_i\left(\mathbf{x}\right)\right)\leq1$, 
\begin{equation*}
    v^{max}_i-\left(v^d_i\tanh\left(-k_{i, 3}g_{v_i}\left(\mathbf{x}\right)d_i\left(\mathbf{x}\right)\right)+v^s_i\right)\geq0.
\end{equation*}
Consequently, the second $\tanh$ function of (\ref{eq.u_tilde_RD1}) is always greater than or equal to 0. If $g_{v_i}\left(\mathbf{x}\right)>0$, then the first $\tanh$ function of (\ref{eq.u_tilde_RD1}) is less than 0, thereby rendering $\dot{h}_{v_i}\left(\mathbf{x}, \mathbf{u}^*\right)\leq0$. If $g_{v_i}\left(\mathbf{x}\right)<0$, then the first $\tanh$ function of (\ref{eq.u_tilde_RD1}) is greater than 0, and thus $\dot{h}_{v_i}\left(\mathbf{x}, \mathbf{u}^*\right)\leq0$.

$\dot{h}_{v_i}\left(\mathbf{x}, \mathbf{u}^*\right)$ being rendered nonpositive for the case when $v_i=v^{min}_i$ could be shown in a similar way. Therefore, $\dot{h}_{v_i}\left(\mathbf{x}, \mathbf{u}^*\right)\leq0$ for all $\mathbf{x}\in\partial\CMcal{S}_{v_i}$, $i\in\left[c\right]$.
\hfill $\blacksquare$ \vspace{0.2cm}


We also emphasize that the nominal evading maneuver $\mathbf{u}^*$ is guaranteed to satisfy the input constraints from (\ref{eq.input_constraint}), as will be shown in the following lemma.
\begin{lemma}
\label{lem.nominal_in_admissible}
The nominal evading maneuver defined as (\ref{eq.nominal_evading_maneuver}) satisfies $\mathbf{u}^*\left(\mathbf{x}\right)\in\CMcal{U}$ for all $\mathbf{x}\in\CMcal{X}$.
\end{lemma}
\noindent \textit{proof.}
For all $i\in\left[m\right]$, the nominal evading maneuver in terms of the modified input given as (\ref{eq.u_tilde_not_RD1}) or (\ref{eq.u_tilde_RD1}) satisfies $\lvert \tilde{u}^*_i\left(\mathbf{x}\right) \rvert <\tilde{u}^{max}_i\left(\mathbf{x}\right)$
for all $\mathbf{x}\in\CMcal{X}$. Since $\tilde{u}^{max}_i\left(\mathbf{x}\right)<\text{min}(\mu_i\left(\mathbf{x}\right),\nu_i\left(\mathbf{x}\right))$
from (\ref{eq.u_tilde_i_max}), $-\mu_i\left(\mathbf{x}\right)<\tilde{u}^*_i\left(\mathbf{x}\right)<\nu_i\left(\mathbf{x}\right)$. Therefore, from (\ref{eq.mu_i_nu_i}) and (\ref{eq.nominal_evading_maneuver}), $u^{min}_i<u^*_i\left(\mathbf{x}\right)<u^{max}_i$, and the nominal evading maneuver satisfies $\mathbf{u}^*\left(\mathbf{x}\right)\in\CMcal{U}$ for all $\mathbf{x}\in\CMcal{X}$.
\hfill $\blacksquare$ \vspace{0.2cm}

\subsection{Ultimate Invariant Set} \label{safety-critical_ultimate}
In this subsection, we define the \textit{ultimate invariant set} $\CMcal{S}_u$ which is a subset of the safe set $\CMcal{S}$ (\ref{eq.safe_set}), and show the existence of an admissible control law $\mathbf{u}:\CMcal{X}\rightarrow\CMcal{U}$ that renders $\CMcal{S}_u$ forward invariant.

The RD2 ZCBF $H_{\mathbf{r}}:\CMcal{X}\rightarrow\mathbb{R}$ is defined using the methodology presented in Theorem \ref{thm.H_ZCBF} with the nominal evading maneuver $\mathbf{u}^*$ from (\ref{eq.nominal_evading_maneuver}):
\begin{equation}
    \label{eq.RD2_ZCBF}
    H_{\mathbf{r}}\left(\mathbf{x}\right):=\sup_{t\geq0}\psi_{h_\mathbf{r}}\left(t;\mathbf{x}, \mathbf{u}^*\right).
\end{equation}
The RD2 ZCBF set $\CMcal{S}_{H_{\mathbf{r}}}$ is defined as the 0-sublevel set of $H_{\mathbf{r}}$: $\CMcal{S}_{H_{\mathbf{r}}}:=\left\{\mathbf{x}\in\CMcal{X}\mid H_{\mathbf{r}}\left(\mathbf{x}\right)\leq 0 \right\}$.
We assume $\frac{\partial H_{\mathbf{r}}}{\partial\mathbf{X}}\left(\mathbf{x}\right)\neq0$ for all $\mathbf{x}\in\partial\CMcal{S}_{H_{\mathbf{r}}}$. 
For later brevity, we will refer to the RD2 ZCBF set $\CMcal{S}_{H_{\mathbf{r}}}$ and the RD1 safe sets $\CMcal{S}_{v_1}, \CMcal{S}_{v_2}, \cdots, \CMcal{S}_{v_c}$ as the \textit{ultimate safe sets}.
We define the ultimate invariant set $\CMcal{S}_u$ as the intersection of all ultimate safe sets:
\begin{equation}
    \CMcal{S}_u = \CMcal{S}_{H_{\mathbf{r}}}\cap\left(\textstyle{\bigcap_{i=1}^{c}}\CMcal{S}_{v_i}\right).
    \label{eq.ultimate_safe_set}
\end{equation}
Note that since the RD2 ZCBF set $\CMcal{S}_{H_{\mathbf{r}}}$ is a subset of the RD2 safe set $\CMcal{S}_{\mathbf{r}}$, the ultimate invariant set $\CMcal{S}_u$ is a subset of the safe set $\CMcal{S}$ (\ref{eq.safe_set}). In other words, all $\mathbf{x}\in\CMcal{S}_u$ obeys the RD2 constraint as well as every RD1 constraint.

We now show that the ultimate invariant set $\CMcal{S}_u$ could be rendered forward invariant by a controller which satisfies the input constraints given as (\ref{eq.input_constraint}).


\begin{theorem}
\label{thm.main}
There exists a controller $\mathbf{u}:\CMcal{X}\rightarrow\CMcal{U}$ that renders the ultimate invariant set $\CMcal{S}_u$ defined as (\ref{eq.ultimate_safe_set}) forward invariant, while satisfying the input constraints given as (\ref{eq.input_constraint}).
\end{theorem}
\noindent \textit{proof.}
We prove the theorem using Nagumo's theorem \cite{blanchini2008set}. Since the ultimate invariant set $\CMcal{S}_u$ is defined as the intersection of multiple ultimate safe sets, $\mathbf{x}\in\partial\CMcal{S}_u$ could be on the boundary of a single ultimate safe set, or where boundaries of multiple ultimate safe sets intersect.

We first consider the case where $\mathbf{x}\in\partial\CMcal{S}_{H_{\mathbf{r}}}$ while $\mathbf{x}\notin\partial\CMcal{S}_{v_i}$ for all $i\in\left[c\right]$. The nominal evading maneuver $\mathbf{u}^*$ defined as (\ref{eq.nominal_evading_maneuver}), which always belongs to $\CMcal{U}$ by Lemma \ref{lem.nominal_in_admissible}, renders $\dot{H}_{\mathbf{r}}\left(\mathbf{x}, \mathbf{u}^*\right)\leq0$ for all $\mathbf{x}\in\CMcal{S}_{H_{\mathbf{r}}}$ \cite{breeden2021high}. In fact, $\dot{H}_{\mathbf{r}}\left(\mathbf{x}, \mathbf{u}^*\right)$ would be rendered nonpositive for all $\mathbf{u}^*:\CMcal{X}\rightarrow{U}$ satisfying $\mathbf{u}^*\in\CMcal{C}^1$. 

We then consider the case where $\mathbf{x}\in\partial\CMcal{S}_u$ while $\mathbf{x}\notin\partial\CMcal{S}_{H_{\mathbf{r}}}$ (i.e., $\mathbf{x}$ is on the boundary of one or more RD1 safe sets). Without loss of generality, we assume $\mathbf{x}\in\partial\CMcal{S}_{v_i}$ for all $i\in\left[k\right]$, where $1\leq k\leq c$. Since the derivative of an RD1 constraint function $\dot{h}_{v_i}\left(\mathbf{x}, \mathbf{u}^*\right)$ depends only on a single input channel $u^*_i$ as can be seen in (\ref{eq.h_v_i_dot}), the nominal evading maneuver $\mathbf{u}^*$ from (\ref{eq.nominal_evading_maneuver}) simultaneously renders $\dot{h}_{v_i}\left(\mathbf{x}, \mathbf{u}^*\right)\leq0$ for all $i\in\left[k\right]$ by Lemma \ref{lem.RD1_safe_set_forward_invariance}. 

Finally, we consider the case where $\mathbf{x}\in\partial\CMcal{S}_{H_{\mathbf{r}}}$ and $\mathbf{x}\in\partial\CMcal{S}_{v_i}$ for all $i\in\left[k\right]$, where $1\leq k\leq c$. The nominal evading maneuver $\mathbf{u}^*$ defined as (\ref{eq.nominal_evading_maneuver}) yields $\dot{H}_{\mathbf{r}}\left(\mathbf{x}, \mathbf{u}^*\right)\leq0$ while simultaneously rendering $\dot{h}_{v_i}\left(\mathbf{x}, \mathbf{u}^*\right)\leq0$ for all $i\in\left[k\right]$. Other forms of nominal evading maneuver that do not take into account the RD1 constraints would still render $\dot{H}_{\mathbf{r}}\left(\mathbf{x}, \mathbf{u}^*\right)\leq0$ as discussed earlier in this proof, but would not be able to guarantee the nonpositiveness of $\dot{h}_{v_i}\left(\mathbf{x}, \mathbf{u}^*\right)$ at the same time.

The above cases show that for all $\mathbf{x}\in\partial\CMcal{S}_u$, the derivatives of all RD2 ZCBF or RD1 constraint functions with zero value at $\mathbf{x}$ are rendered nonpositive by at least one admissible control law $\mathbf{u}:\CMcal{X}\rightarrow\CMcal{U}$, namely the nominal evading maneuver $\mathbf{u}^*$.
Furthermore, $\frac{\partial h_{v_i}}{\partial\mathbf{x}}\left(\mathbf{x} \right )\neq0$ for all $\mathbf{x}\in\partial\CMcal{S}_{v_i}$ and $i\in\left[c\right]$, and $\frac{\partial H_{\mathbf{r}}}{\partial\mathbf{x}}\left(\mathbf{x} \right )$ is assumed to be nonzero for all $\mathbf{x}\in\partial\CMcal{S}_{H_{\mathbf{r}}}$.
Therefore, by Nagumo's theorem, the ultimate invariant set $\CMcal{S}_u$ could be rendered forward invariant by a controller that satisfies the input constraints given as (\ref{eq.input_constraint}).
\hfill $\blacksquare$ 
%
\begin{remark}
It is crucial to compute a sufficiently large ultimate invariant set $\CMcal{S}_u$ inside the safe set $\CMcal{S}$ \cite{breeden2021high}. 
Unlike the prior works that design multiple ZCBFs from the state constraints \cite{xu2018constrained, cortez2022robust, cortez2022safe, breeden2022compositions}, we did not construct additional ZCBFs for the RD1 constraints. 
Instead, in Theorem \ref{thm.main}, we have shown that the intersection of the RD2 ZCBF set and the RD1 safe sets could be rendered forward invariant with an admissible control law.
This enables us to fully utilize the RD1 safe sets, not the ZCBF sets that are subsets of the RD1 safe sets, in the construction of $\CMcal{S}_u$, which results in a less conservative invariant set.
\end{remark}

\begin{remark}
Since the greedy control law $\mathbf{u}^*_{\textup{greedy}}$ from (\ref{eq.u_ball}) pointwise minimizes $\ddot{h}_{\mathbf{r}}$ and therefore attempts to minimize $h_{\mathbf{r}}$, it may help to obtain a less conservative RD2 ZCBF set $\CMcal{S}_{H_{\mathbf{r}}}$ when compared to other designs of nominal evading maneuver that does not minimize $\ddot{h}_{\mathbf{r}}$. Therefore, the nominal evading maneuver $\mathbf{u}^*$ from (\ref{eq.nominal_evading_maneuver}), which is specifically designed to mimic $\mathbf{u}^*_{\textup{greedy}}$, may as well result in a less conservative ultimate invariant set $\CMcal{S}_u$.
\end{remark}

\subsection{Safety-Critical One-Step MPC} \label{safety-critical_MPC}
Here, we present a safety-critical controller with guaranteed feasibility that utilizes the results from Theorem \ref{thm.main}. Similar to existing safety-critical controllers, we first impose the RD2 constraint in the form of ZCBF constraint (\ref{eq.ZCBF_input}). On the contrary, since we did not design ZCBFs for the RD1 constraints to obtain a less conservative invariant set, RD1 constraints cannot be written as input affine constraints (\ref{eq.ZCBF_input}). Instead, they should be formulated as $h_{v_i}\left(\mathbf{v}\right)\leq0$ for all $i\in\left[c\right]$, which are not applicable to a QP.
This motivates the formulation of a safety-critical one-step MPC that constrains the system to be inside the ultimate invariant set $\CMcal{S}_u$:
\begin{subequations}
\label{eq.nlp}
\begin{align}
\mathbf{u}_{t, \textup{safe}}=\argminB_{\mathbf{u}_t\in\CMcal{U}}\hspace{0.1cm}&\left \| \mathbf{u}_t-\hat{\mathbf{u}}_t \right \|^2_{R_1} +\left \| \mathbf{u}_t-\mathbf{u}_{t-1} \right \|^2_{R_2},& \label{eq.nlp_cost}\\
\textup{s.t.}\hspace{0.1cm}&\hspace{0.1cm}\mathbf{x}_{t+1}=F\left(\mathbf{x}_t, \mathbf{u}_t \right ),& \label{eq.nlp_discretized_dynamics}\\
&\hspace{0.1cm}\dot{H}_{\mathbf{r}}\left(\mathbf{x}_t, \mathbf{u}_t\right)\leq\alpha\left(-H_{\mathbf{r}}\left(\mathbf{x}_t\right)\right),& \label{eq.nlp_discretized_RD2_ZCBF}\\
&\hspace{0.1cm}h_{v_i}\left(\mathbf{v}_{t+1}\right)\leq0, \forall i\in\left[c\right],\label{eq.nlp_discretized_RD1_constraint}
\end{align}
\end{subequations}
where subscript $t$ denotes the value of the related vector at time $t$.
The first term of the cost function (\ref{eq.nlp_cost}) is similar to that of CBF-QP \cite{ames2016control} except that it is now weighted by a positive definite matrix $R_1\in\mathbb{R}^{m\times m}$, where $\hat{\mathbf{u}}:\CMcal{X}\rightarrow\CMcal{U}$ is a nominal feedback controller that achieves some performance goal (e.g., trajectory tracking), but without consideration of state constraints. The second term of (\ref{eq.nlp_cost}), which is weighted by a positive semi-definite matrix $R_2\in\mathbb{R}^{m\times m}$, is added to decrease chattering, and does not hinder safety guarantee. The condition imposed on $\mathbf{u}_t$ in (\ref{eq.nlp_cost}) represents the input constraints from (\ref{eq.input_constraint}), and (\ref{eq.nlp_discretized_dynamics}) describes the system dynamics (\ref{eq.general_system}) in the discrete-time domain. (\ref{eq.nlp_discretized_RD2_ZCBF}) represents the RD2 ZCBF constraint (\ref{eq.ZCBF_input}), and the RD1 constraints are implemented as (\ref{eq.nlp_discretized_RD1_constraint}). Together, (\ref{eq.nlp_discretized_RD2_ZCBF}) and (\ref{eq.nlp_discretized_RD1_constraint}) assure the system state to remain inside the ultimate safe set $\CMcal{S}_u$.

\begin{proposition}
\label{propo.nlp_feasibility}
The safety-critical one-step MPC (\ref{eq.nlp}) is always feasible, given $\mathbf{x}_t\in\CMcal{S}_u$.
\end{proposition}
\noindent \textit{proof.}
The nominal evading maneuver $\mathbf{u}^*$ from (\ref{eq.nominal_evading_maneuver}) renders $\dot{H}_{\mathbf{r}}\left(\mathbf{x}_t, \mathbf{u}^*\right)\leq0\leq\alpha\left(-H_{\mathbf{r}}\left(\mathbf{x}_t\right)\right)$ for all $\mathbf{x}_t\in\CMcal{S}_u\subset\CMcal{S}_{H_{\mathbf{r}}}$ \cite{breeden2021high}. Moreover, as shown in the proof of Theorem \ref{thm.main}, the derivatives of all RD1 constraint functions with zero value at $\mathbf{x}_t$ are rendered nonpositive by $\mathbf{u}^*$, and thus assures $h_{v_i}\left(\mathbf{v}_{t+1}\right)\leq0$ for all $i\in\left[c\right]$. Therefore, at least one admissible control input, namely $\mathbf{u}^*$, simultaneously satisfies the constraints (\ref{eq.nlp_discretized_RD2_ZCBF}) and (\ref{eq.nlp_discretized_RD1_constraint}).
\hfill $\blacksquare$ \vspace{0.2cm}

The proposed framework is guaranteed to be recursively feasible since the controller (\ref{eq.nlp}) is feasible given $\mathbf{x}_t\in\CMcal{S}_u$ and the resulting optimal input yields $\mathbf{x}_{t+1}\in\CMcal{S}_u$. Thus, we formulate a one-step MPC to enhance real-time applicability.

\section{Application to Fixed-Wing UAV}
In this section, we demonstrate the safety-critical controller (\ref{eq.nlp}) in simulation on a fixed-wing UAV.

\subsection{Dynamical Model and Constraints}
The dynamical model of a fixed-wing UAV is given as follows \cite{lee2011obstacle, beard2012small}:
\begin{equation}
    \begin{bmatrix}
    \dot{P_x}\\ 
    \dot{P_y}\\ 
    \dot{P_z}\\
    \dot{V}\\ 
    \dot{\gamma}\\ 
    \dot{\psi}
    \end{bmatrix}=\begin{bmatrix}
    V\cos\gamma\cos\psi\\ 
    V\cos\gamma\sin\psi\\ 
    V\sin\gamma\\
    u_V\\ 
    \left(u_\gamma-g\cos\gamma\right)/V\\ 
    u_\psi/\left(V\cos\gamma\right)
    \end{bmatrix},
    \label{eq.UAV_dynamics}
\end{equation}
where the system state and control input are defined as $\mathbf{x}=\left[P_x, P_y, P_z, V, \gamma, \psi\right]^\top\in\mathbb{R}^6$ and $\mathbf{u}=\left[u_V, u_\gamma, u_\psi\right]^\top\in\mathbb{R}^3$. 
The RD2 states $\mathbf{P}:=\left[P_x, P_y, P_z\right]^\top$ represent the position of the UAV in the inertial coordinate frame.
Flight speed, flight path angle, and heading angle of the UAV, which are denoted $V$, $\gamma$, and $\psi$, correspond to the RD1 states. 
Note that (\ref{eq.UAV_dynamics}) could be written in the form of (\ref{eq.general_system}).

The UAV is subject to multiple safety constraints. 
First, we consider an RD2 constraint function $h_{obs}:\mathbb{R}^3\rightarrow\mathbb{R}$ which is designed for the task of avoiding a spherical obstacle:
\begin{equation}
    h_{obs}\left(\mathbf{P}\right)=\left(R+R_{obs}+R_{min}\right)^2-\left \| \mathbf{P} - \mathbf{P}_{obs}  \right \|^2,
\end{equation}
where $R$, $R_{obs}$, and $R_{min}$ represent radius of the sphere that circumscribes the UAV, radius of the spherical obstacle, and the minimum distance to be maintained between the UAV and the obstacle. $\mathbf{P}_{obs}$ is the position of the center of the obstacle. The corresponding RD2 ZCBF is obtained from (\ref{eq.RD2_ZCBF}).
Next, the RD1 states $V$ and $\gamma$ should be bounded in the form of box constraints in order to prevent aggressive maneuvers and aerodynamic stall \cite{khare2022predictive, bayen2007aircraft}. Such constraints are formulated using the RD1 constraint function (\ref{eq.RD1_constraint_function}).
Finally, the system (\ref{eq.UAV_dynamics}) is under input constraints that arise from the actuator limits of real-world dynamical systems. 
We represent the set of admissible inputs $\CMcal{U}$ in the form of (\ref{eq.input_constraint}).

\subsection{Simulation Results}
\begin{figure}
    \centering
    \includegraphics[width=1.0\linewidth,trim={1.5cm 2.1cm 1cm 2.6cm},clip]{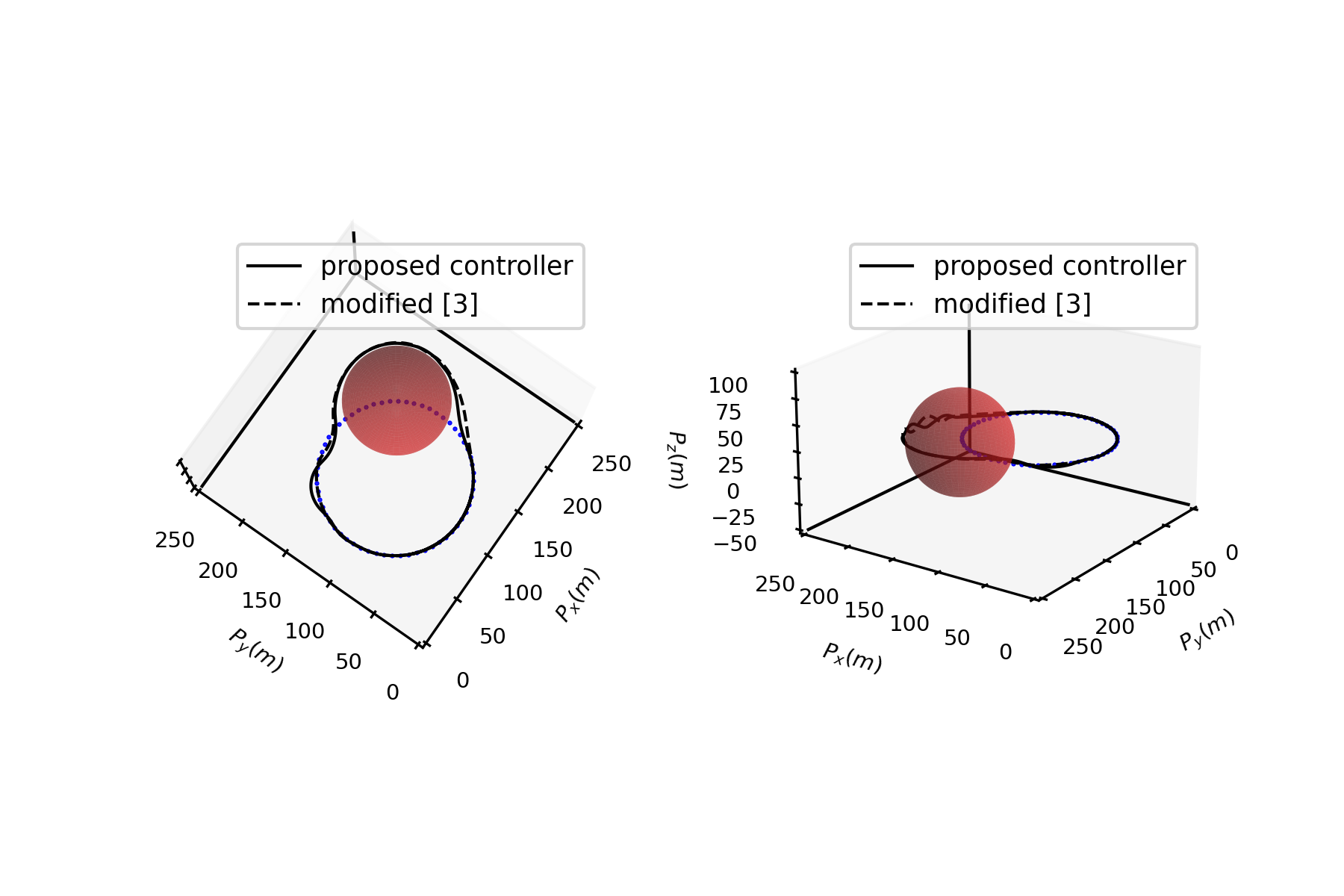}
    \caption{Position trajectories traversed during the operation time of 70 seconds. The left figure shows the top-down view, and the right figure depicts the oblique view. The red sphere represents the obstacle to avoid.
    }
    \label{fig.trajectories}
    \vspace{-0.5cm}
\end{figure}

As for the nominal feedback controller $\hat{\mathbf{u}}$, we utilize an MPC that tracks a circular reference trajectory. 
To avoid confusion with the safety-critical one-step MPC, we will refer to the nominal feedback controller as the nominal MPC. 
The control horizon of the nominal MPC is set as 0.5 \si{s}, and its cost function is formulated as a sum of quadratic terms that are designed for trajectory tracking and input regulation. 
The nominal MPC takes into account the input constraints, however the RD2 and RD1 constraints are not considered. 
To demonstrate the validity of the proposed safety-critical controller, a circular reference trajectory that penetrates the obstacle as can be seen in Fig. \ref{fig.trajectories} is provided. Collision is unavoidable without an additional safety-critical controller.

We compare two types of safety-critical controllers: the proposed safety-critical one-step MPC and the ZCBF-based controller presented in \cite{breeden2021high}. The two resultant features of the proposed controller compared to the other are that 1) additional RD1 constraints can be satisfied and 2) input chattering can be reduced.
In implementing the ZCBF-based controller, owing to the originally adopted nominal evading maneuver in \cite{breeden2021high} not being continuously differentiable, we slightly modify the method by using the continuously differentiable nominal evading maneuver (\ref{eq.u_tilde_not_RD1}) instead. Such controller will be referred to as ``modified \cite{breeden2021high}" in the following figures.

\begin{figure}
    \centering
    \includegraphics[width=1.0\linewidth,trim={0.4cm 1.2cm 0.4cm 2cm},clip]{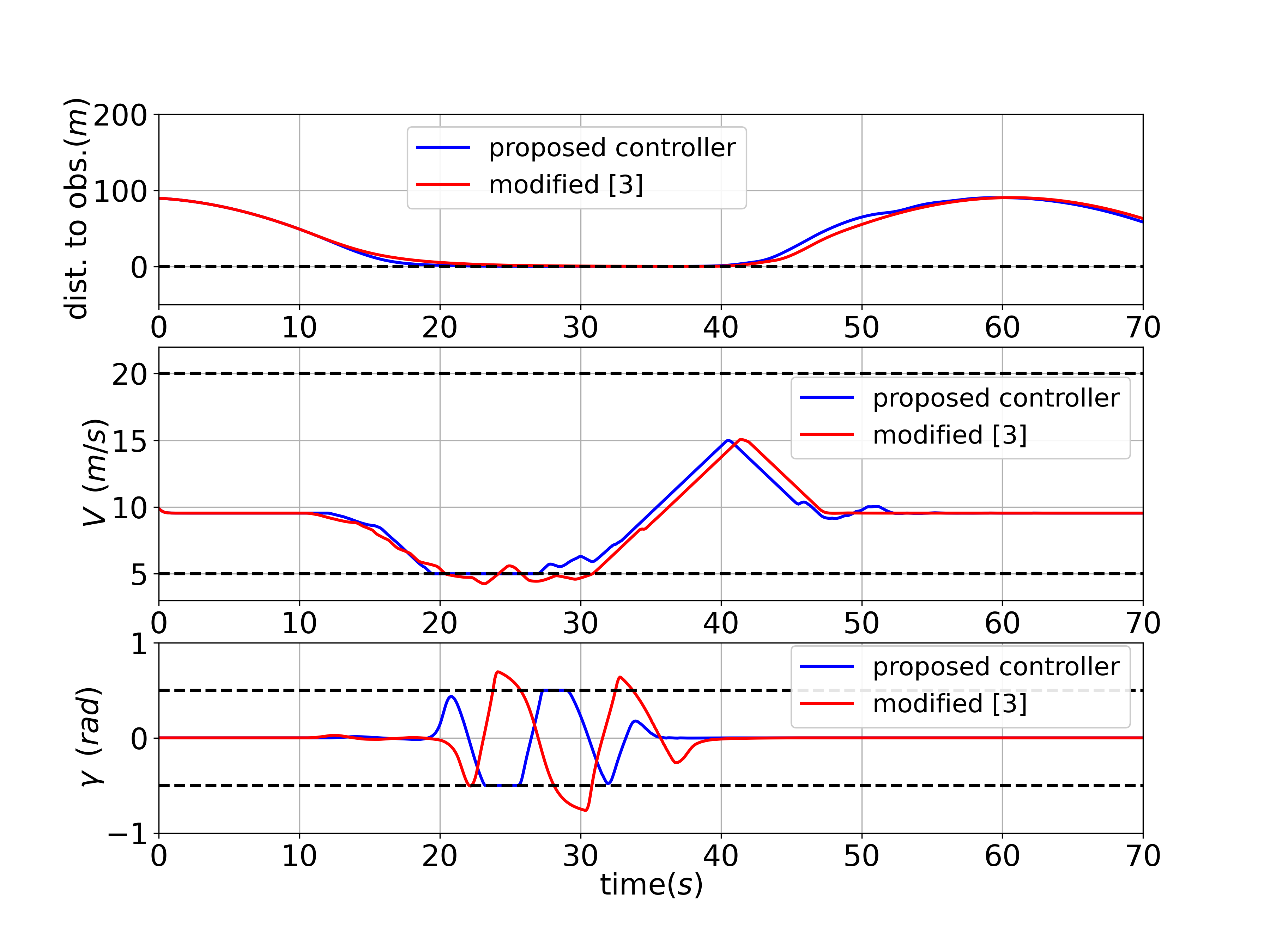}
    \caption{From top to bottom, the figures plot the distance between the UAV and the obstacle, flight speed $V$, and flight path angle $\gamma$ against time.}
    \label{fig.states}
\end{figure}

\begin{figure}
    \centering
    \begin{subfigure}{1.0\linewidth}
        \centering
        \includegraphics[width=1\linewidth,trim={0cm 0.28cm 0cm 0.2cm},clip]{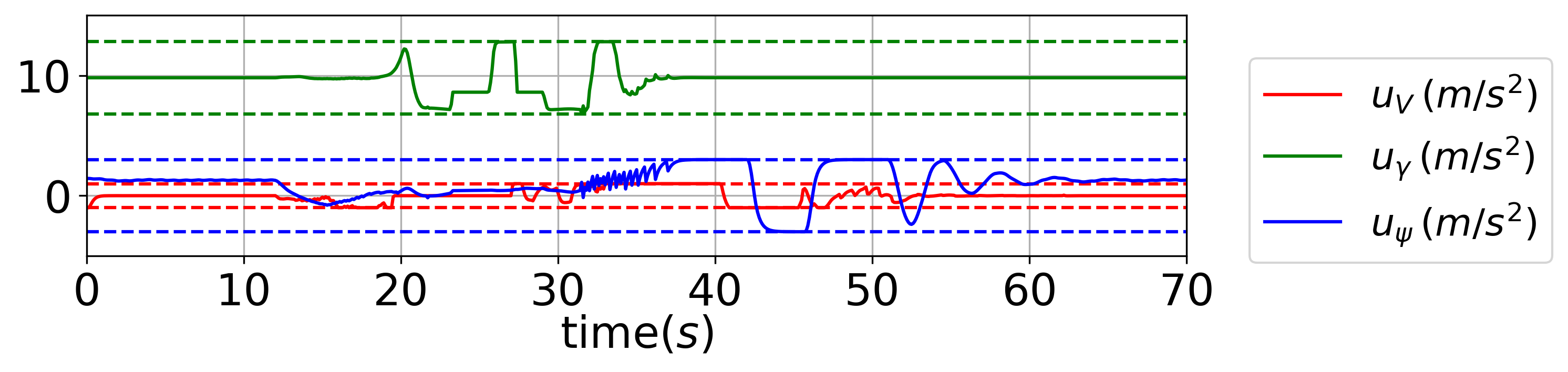}
        \caption{Control inputs from the proposed controller (\ref{eq.nlp}).}
        \label{fig.input_proposed}
    \end{subfigure}
    \hfill
    \begin{subfigure}{1.0\linewidth}
        \centering
        \includegraphics[width=1\linewidth,trim={0cm 0.28cm 0cm 0cm},clip]{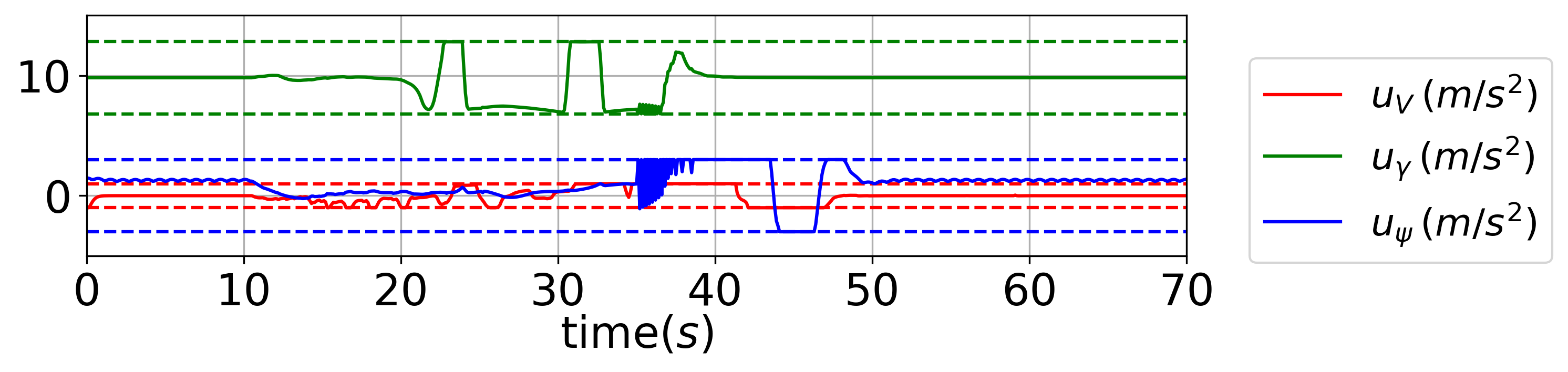}
        \caption{Control inputs from modified \cite{breeden2021high}.}
        \label{fig.input_breeden}
    \end{subfigure}
    \caption{Time history of control inputs. While both controllers satisfy the input constraints, less input chattering can be obtained with the proposed controller (\ref{eq.nlp}).}
    \label{fig.input}
    \vspace{-0.5cm}
\end{figure}

The simulation results of the UAV using either the proposed safety-critical one-step MPC (\ref{eq.nlp}) or the modified version of \cite{breeden2021high} as the controller are shown in Figs. \ref{fig.states}, \ref{fig.input}, and \ref{fig.trajectories}\footnote{Implementation details including the parameteres and codes can be found at \texttt{https://github.com/DonggeonDavidOh/LARR2023.git}}. As can be seen in Fig. \ref{fig.trajectories}, both types of controllers were able to guarantee safety regarding the RD2 constraint which represents collision avoidance. The same conclusion can be made from Fig. \ref{fig.states}, where the distance between the UAV and the obstacle was kept positive for both types of controllers. 

However, notice that the modified version of \cite{breeden2021high} was unable to restrict the RD1 states $V$ and $\gamma$ between their maximum and minimum allowable values, which are represented as black dotted lines in Fig. \ref{fig.states}. In contrast, the proposed one-step MPC successfully bounded $V$ and $\gamma$ to remain inside the corresponding RD1 safe sets. Moreover, the proposed controller was able to find a safe and feasible control input even when the UAV was simultaneously on the boundary of multiple ultimate safe sets, which is observable in Fig. \ref{fig.states} at $t=23 \ \si{s}$. Therefore, we conclude that the proposed controller successfully handles the case of overlapping constraint functions, where boundaries of multiple safe sets intersect.

We also state that although both types of controllers have generated control inputs that strictly satisfy the input constraints plotted as dotted lines in Fig. \ref{fig.input}, the modified version of \cite{breeden2021high} suffered from chattering. The one-step MPC eases the issue by introducing the second term in (\ref{eq.nlp_cost}).

Finally, we stress the real-time applicability of the proposed one-step MPC. The average computation time of the controller was 35.9 \si{ms} with the standard deviation of 1.72 \si{ms}. All codes were written in python3, and the simulation was executed on a laptop equipped with an Intel\textsuperscript{\textregistered} Core\textsuperscript{TM} i7-11800H and 32GB DDR4 RAM which runs the Ubuntu 20.04 operating system.


\section{Conclusion}

We proposed a framework to guarantee safety in the presence of multiple state and input constraints for a class of second-order nonlinear systems. 
The presented framework is able to handle three types of constraints: a constraint that restricts states of relative degree 2 (RD2 constraint), box constraints for states of relative degree 1 (RD1 constraints), and input constraints, all of which could be used to represent safety-critical situations including collision avoidance, velocity bounds, and input saturation.
First, we devised the RD2 ZCBF, a zeroing control barrier function induced from the RD2 constraint, using a new nominal evading maneuver. Since the nominal evading maneuver was carefully designed to 1) be continuously differentiable, 2) satisfy input constraints, and 3) be capable of handling other RD1 constraints, we were able to prove that the ultimate invariant set, which is defined as the intersection of the RD2 ZCBF set and RD1 safe sets, could be controlled invariant. Then, the ultimate invariant set was utilized to construct a safety-critical one-step MPC with guaranteed recursive feasibility. By virtue of its one-step nature, the proposed controller is computationally tractable and adoptable in real-time applications. We demonstrated the effectiveness of the safety-critical controller in simulation where a fixed-wing UAV tracks a circular trajectory while strictly abiding by multiple constraints, which are collision avoidance constraint, bounds on flight speed and flight path angle, and input constraints.

\normalem 

\end{document}